\documentstyle[aps,preprint]{revtex}
\begin{document}
\draft
\title {Square to stripe transition and superlattice patterns 
in vertically oscillated  granular layers}
\author{Hwa-Kyun Park\footnote{e-mail: childend@complex.kaist.ac.kr \\
 TEL:+49-351-871-1225 \hspace{.5cm} FAX: +49-351-871-1999}}
\address{ Max-Planck Institut f\"ur Physik komplexer Systeme, 
Dresden 01187, Germany } 
\author{Hie-Tae Moon.} 
\address{Department of Physics,Korea Advanced Institute of Science and
Technology,Taejon 305-701,Korea} 
\date{12 November 2001}

\maketitle

\begin{abstract}
We investigated the physical mechanism for the pattern transition from
square lattice to stripes, which appears in vertically oscillating
granular layers. We present a continuum model to show that 
the transition depends on the competition between inertial force 
and local saturation of transport.
By introducing multiple free-flight times, this model further
enables us to analyze the formation of superlattices as well 
as hexagonal lattices 
\end{abstract}


Vertically oscillated granular layers exhibit various interesting
patterns such as subharmonic stripes, squares, hexagons as well as 
localized structures called oscillons
\cite{mus94,mus95,mus96,ums98,mkj97,us00,Pak01}.
The control parameters are the driving frequency $f$
and the dimensionless acceleration $\Gamma=4 \pi ^{2} f^{2} A/g$,
where $g$ is the acceleration of gravity, and $A$ is the amplitude of
the vibrating surface.
Increasing $\Gamma$ with fixed $f$, flat states, subharmonic squares
or stripes depending on $f$, subharmonic hexagons and flat states
appear in sequence.

These granular patterns have been studied using 
phenomenological models
\cite{s97,cmr97,ta97,at98,vo98,rothman,jm99,eggers} 
and molecular dynamics simulations\cite{bs98}.
The studies show that transitions between granular patterns 
as $\Gamma$ increases are related to the period doubling
bifurcation\cite{mus95,vo98}.
But the bifurcation as a function of $f$, that is the transition
between squares and stripes, has not been well understood yet.
There have been many models showing both squares and 
stripes \cite{s97,ta97,vo98,bs98} and
it was suggested that the effective viscosity and the horizontal
mobility resolve the transition\cite{mus95,s97,cmr97}.
Experiments also show that squares arise at large lateral transport 
while stripes form at small lateral transport\cite{mkj97}.
However, there has been no explicit model for the  
square to stripe transition upon increasing frequency.

In this paper, we propose a continuum model which describes the
transition between squares and stripes. We incorporate
the lateral transfer and the local saturation mechanism
into our model. The study shows that the ratio of these two
competing processes depends on the frequency, which explains
the dependence of the square-stripe transition on
the drive frequency.
In addition, considering multiple free-flight times,
we study the hexagonal and superlattice patterns which 
are observed in the granular system with multiple 
frequency forcing \cite{Pak01}.

The granular patterns are similar to Faraday waves in vibrated fluids
\cite{kg97}, however some differences exist.
\begin{itemize}
\item[(i)]  On each cycle, there is a free-flight time during which
granular layer loses contact with a bottom plate. 
\item[(ii)] In general, macroscopic physical quantities
for the granular material are not well defined.
But Ref. \cite{us00} shows that a continuum description of the 
granular layer is valid for the square-stripe transition. 
Molecular simulation results \cite{bs98} also suggest that 
granular continuum equations can be used as far as 
pattern formation of vertically oscillated granular media is
concerned.
\item[(iii)] Granular material has an angle of repose, and
shows both solid-like and liquid-like motions. 
\end{itemize}

Considering (i),(ii), a continuum equation is introduced with
two steps: contact with bottom plate, and free-flight. 
It provides the two competing processes, diffusion during the 
contact and focusing effect during the
free flight, which were suggested as mechanisms of subharmonic wave
generation\cite{cmr97}.
A hysteresis effect due to (iii), which accounts for the
formation of oscillon\cite{cmr97,ta97,vo98,rothman,jm99,eggers}, 
is not considered here since it appears 
to be unrelated to the square-stripe transition\cite{us00}.

At $t=0$, the layer collides with the bottom plate and the slope
of granular layers begins to relax. From the collision, granular particles
get kinetic energy and move randomly. The process can be described
as diffusion, and the following equation is obtained for
$0 \leq t<t_{c}$ during which granular layers have contact with 
the bottom plate
\begin{equation} \label{diffusion}
    \rho_t=D \nabla ^{2} \rho .
\end{equation}
We assume that the thickness h is a monotonic increasing
function of the local density $\rho$, where $\rho$ is a mass of
granular material per unit area, and choose $\rho$ as 
an order parameter.

At $t=t_{c}$, the layer loses contact with 
the bottom plate. During the free flight, the lateral 
transfer of grains induce the focusing of the sands, 
so that subharmonic excitation of the granular wave occurs. 
From continuity,  the flux $\vec{j}$ and the velocities 
$\vec{v}$ of grains right after takeoff from the bottom 
plate are given by

\begin{equation}\label{jv}
\begin{array}{c}
   \vec{j}(\vec{x},t_{c})=-D \nabla \rho(\vec{x},t_{c})
\\
\\ \vec{v}(\vec{x},t_{c})=\vec{j}(\vec{x},t_{c})/ \rho(\vec{x},t_{c})
    = -D \nabla \rho(\vec{x},t_{c}) / \rho(\vec{x},t_{c}) .
\end{array}
\end{equation}

This is indeed what was done by Cerda {\it et. al.}. \cite{cmr97}, in
which saturation of the flux is not considered, and only square
patterns arise. Hence one may expect that the saturation is
related to the formation of stripe patterns.
Assuming that granular layers behave like fluid during the free 
flight, we adopt Navier-Stokes equations in 2 dimensions as 
a model of the the local saturation of flux as well as
the transfer of particles and momentum.

During the interval $t_{c} \leq t< T$, where $T=1/f$ is the period of
the driving force, we get 

\begin{equation} \label{nse}
\begin{array}{c}
    \left\{ \begin{array}{c}
         \rho_{t}+ \nabla \cdot (\rho \vec{v})=0  \\
	   \\
         \vec{v}_{t}+\vec v \cdot \nabla \vec v =
         -\frac{1}{\rho} \nabla p.
           \end{array} \right.   
\end{array}
\end{equation}

Momentum is transferred through the transport of the granular 
particle, and inelastic collisions between the particles suppress 
the lateral motions.
In general, these effects can be incorporated into the local stress
tensor $T_{ij}$. 
Assuming that the suppression of the lateral motion is most 
important, the local stress tensor is approximated
by $T_{ij}= -p \delta_{ij}.$
When we consider the collision between granular particles, the
relevant variable is a relative velocity rather than a velocity itself.
A dimensional consideration leads to the following form of
the pressure
\begin{equation}\label{p}
    p= - \alpha \rho |\nabla \cdot \vec v|
    \nabla \cdot \vec v .
\end{equation}
The pressure is high at the accumulation points
where $\nabla \cdot \vec v < 0$ , and
low at the dispersion points where $\nabla \cdot \vec v > 0$.
The pressure difference produces the force to
suppress the focusing of sands.
The parameters of Eq. (\ref{diffusion}),(\ref{jv}), (\ref{nse}) 
and (\ref{p}) are $D,f,\alpha$ and $\tau=t_{f}/T=1-t_{c}/T$
where $t_{f}$ is a free flight time.
In a real system, $\tau$ is determined mainly by $\Gamma$.
$D$ and $\alpha$ depend on layer depth, grain diameter,
and restitution coefficient as well as $\Gamma$ and $f$.
We choose $\alpha=0.015, D=0.17$, and system size $l=10$.

Linear stability analysis for the rest state shows that 
the amplitude of the mode with wave number $k$ is amplified 
by the Floquet multiplier
$F(k)=(1-D k^2 \tau T) \exp (-D k^2 (1-\tau) T)$ on each cycle.
At small $\tau$, we get only flat states.
As $\tau$ is increased beyond $\tau_c=0.782$, which is 
independent of $D$ and $T$, minimum value of
$F(k)$ becomes less than -1 and subharmonic standing waves arise.  
In this case, we obtain the dispersion relation,
$f=D \tau (1-\tau) k^2$.
Note that the value of $\tau_c$ does not depend on $D,f,\alpha$ 
and $l$.
The bifurcation diagram of Eq. (\ref{diffusion})
and (\ref{nse}) as a function of $f$ and $\tau$ is presented
in Fig.\ref{phase}; Fig.\ref{trans} describes typical patterns.
With increasing f with fixed $\alpha$ and $\tau$ , one can see the square
to stripe transition as in Fig.\ref{trans} (a),(b).

Increasing $\alpha$ with fixed f and $\tau$ also
yields the square to stripe transition. 
In other words, the local saturation prefers stripe patterns.
It is interesting to compare the results with the Faraday waves.
In Faraday waves, squares are observed for small viscosity while
stripes for large viscosity\cite{kg97}. In this case viscosity
represents the strength of local saturation which competes with
the inertial force.

To understand the relation between the square-stripe transition 
and the drive frequency, let us write the equations in dimensionless
forms.  Naturally, we take $T=1/f$ as the time unit. For the 
length unit, we choose $\sqrt{D/f}$ which reflects the wavelength.
Inserting $t=\hat t/f $, $\nabla=\hat \nabla / \sqrt{D/f}$,
$v=\sqrt{Df} \hat v$, into Eq. (\ref{diffusion}),
Eq. (\ref{nse}) with Eq. (\ref{p}), and dropping the carets,
we get for $0 \leq t< 1-\tau$,

\begin{equation} \label{diffusion2}
    \rho_t= \nabla ^{2} \rho 
\end{equation}
and for $1 -\tau \leq t < 1$,
\begin{equation} \label{nse2}
\begin{array}{c}
    \left\{ \begin{array}{c}
         \rho_{t}+ \nabla \cdot (\rho \vec{v})=0  \\
	   \\
         \vec{v}_{t}+\vec{v} \cdot \nabla \vec v =
         \frac{R}{\rho} \nabla
		 ( \rho |\nabla \cdot \vec{v}| \nabla \cdot \vec{v})
           \end{array} \right.   
\end{array}
\end{equation}
where $R$ is defined as $\frac{f \alpha}{D}$.

Now we have two dimensionless parameters. One is $\tau=t_{f}/T$, 
the fraction of flight time for one period, and the other is
$R=\frac{f \alpha}{D}$ in Eq. (\ref{nse2}),
which corresponds to the ratio of the force causing the local
saturation to the inertial force.
Note that $R$ is proportional to the frequency $f$. It means that
increase in $f$ induces the strong local saturation
compared to the lateral movement. 
Qualitatively, this can be explained as follows.
Lateral movements are suppressed by the inelastic collisions
which are reflected by pressure $p$ in our model. 
When wavelengths are short, the velocity differences between the 
neighbor grains are large, then so is pressure $p$ from Eq. (\ref{p}).
Hence, for high frequencies, the collision is dominant and mobility 
is suppressed. 

Experimentally, the dispersion relation has a kink structure, i.e., it  
takes a different form below and above $f=f_c$ \cite{mkj97,us00,bs98}. 
For almost all conditions, the square-stripe transition occurs
for $f<f_c$ \cite{us00}. 
In other words, only the dispersion relation for $f<f_c$ is 
relevant as far as the square-stripe transition is concerned.
In this regime, earlier experiments \cite{mus94,mkj97} reported
an exponent 2, however a recent experiment \cite{us00} yielded 
a different exponent value of 1.3.
In contrast, in our model, the dispersion relation 
$\lambda \sim (D/f)^{1/2}$ is obtained. This contradiction 
can be resolved by considering the frequency dependency
of the diffusion coefficient $D$ \cite{cmr97,eggers}.
Considering the relation between viscous length and 
energy injection rates, for fixed $\Gamma$, $D$ scales 
like $1/f^3$ in the small frequency regime \cite{cmr97,lcb94}.
With this, $\lambda \sim (1/f^{2})$ is obtained and
the parameter $R$ which measures the ratio of the saturation of
mobility to the inertial force is proportional to $f^4$. 
Also note that influences of $\Gamma$ on $D$ and $\alpha$ 
have not been considered. Dependence of the square to stripe transition
frequency on $\tau$ in Fig.\ref{phase}, in contrast to the experimental result
\cite{us00} where the transition frequency depends only on the layer depth 
$H$, may be attributed to this.
Nevertheless, for fixed $\Gamma$, the above conclusion that squares transit 
to stripes with increasing the drive frequency does not change.

Next we consider the instability of stripe patterns. With an abrupt
change of the driving frequency $f$, we observe a crossroll 
and a zigzag instability.
Fig. \ref{stability}(a) shows an example of the crossroll
instability.
The stripes are unstable to crossrolls on both the high and low $k$
sides in our model. Fig. \ref{stability}(b) illustrates
the increase in wave vector k due to the zigzag instability.
In experiments and molecular dynamics simulations\cite{bsg98},
the stripe patterns have the crossroll instability
for decreasing $k$, and the skew-varicose instability for increasing $k$. 
The zigzag instability has not been observed in experiments.
Refining the stress tensor and comparing the results with
the instabilities in experiments will yield more insights about
the interaction between granular particles.

As external forcing $\Gamma$ is increased,
the motions are repeated after two different flight
times and contact times.
This period doubling causes hexagonal patterns \cite{mus95},
and the mechanism is investigated theoretically \cite{vo98}.
Recently, it has been reported that multiple frequency forcing on the
granular layer yields superlattice patterns \cite{Pak01}. 
Similar patterns have been studied in Faraday waves
\cite{Faraday}, driven ferrofluid surfaces \cite{Ferro}, 
nonlinear optics \cite{Optics}, Turing patterns \cite{Turing}
and Rayleigh-Benard convection \cite{Rayleigh}. As in the 
single frequency forcing \cite{mus96}, the important variables
determining patterns are expected to be the free-flight times 
and contact times. 

To study the hexagons and superlattice patterns,
we introduce the multiple free-flight times
$t_{f1},t_{f2},...,t_{fn}$ and contact times 
$t_{c1},t_{c2},...,t_{cn}$. 
Let's apply Eq.(\ref{diffusion}) during $t_{c1}$ and Eq.(\ref{nse}) 
during $t_{f1}$. After that, we apply again Eq.(\ref{diffusion})
during $t_{c2}$, Eq.(\ref{nse}) during $t_{f2}$ and so on.
The case with $n=2$ reduces to the period doubling 
and hexagonal patterns are obtained.

Linear stability analysis shows that the amplitude 
of the mode with wave number $k$ is amplified by the factor 
$F(k)=\prod_{i=1}^{n} (1-D k^2 t_{fi}) 
\exp (-D k^2 t_{ci})$ on each cycle.
For the hexagonal patterns, as expected, $F(k)$ gives
a harmonic mode as in Fig. \ref{super}(a). 
However, in general, $F(k)$ can yield instabilities
with various wavelengths and, as a result, 
superlattice patterns can be formed. 
In Fig. \ref{super}(b),
the modes with $\vec{k_1}$ and $\vec{k_2}$ are subharmonic 
and the mode $\vec{q}$ is harmonic. They satisfy
the resonance condition $\vec{k_1}+\vec{k_2}+ \vec{q}=0$. 
Evidently, there is a large freedom of choice for the 
values of $t_{ci}$ and $t_{fi}$. Analysing the characteristics 
of the superlattice patterns in these vast parameter 
spaces is a problem for future studies.

In conclusion, based on the mass and momentum conservation law,
we presented the continuum model to explain the
transition between squares and stripes in vertically oscillated
granular layers.
Patterns are selected by two competing nonlinear interaction:
the inertial force and the local saturation due
to the inelastic collisions between granular particles.
With increasing the drive frequency, the local saturation gets
stronger compared to the inertial force effect, so that
the square to stripe transition occurs.
Introducing multiple free-flight times and contact times yields
hexagonal patterns and superlattice patterns. 

We thank S.-O.Jeong, T.-W.Ko and P.-J. Kim for useful 
discussions, M. B\"aer for useful comments,
and H.K. Pak and P. Umbanhowar for notifying us of
their experimental results.
This work was supported by the interdisciplinary research
program of the KOSEF(Grant No. KRF-2000-015-DP0097).

\pagebreak
\newpage


\begin{figure}
\caption{
	Bifurcation diagram as a function of $f$ and $\tau$.
	$\alpha=0.015,D=0.17,l=10$
    }
\label{phase}
\end{figure}

\begin{figure}
\caption{ 
Square and stripe patterns. 
White corresponds to the large $\rho$.
$D=0.17$,$l=10$, $\alpha=0.015$, $\tau=0.85$.  
(a) $T=1.2$, (b) $T=0.7 $.}
\label{trans}
\end{figure}

\begin{figure}
\caption{ Instabilities of stripes.
$\alpha=0.015,D=0.17,l=10$. For $t<0$, $T=0.55$, and at t=0,
$T=1/f$ is changed abruptly.
(a) T=0.85, Crossroll instability.
(b) T=0.30, Zigzag instability.
}
\label{stability}
\end{figure}

\begin{figure}
\caption{ Hexagon and superlattice patterns.
Left : Hexagonal and superlattice patterns; 
Middle : Power spectra of the left images;
Right : Floquet multiplier $F(k)$ vs $k$.    
$D=0.17,l=10, \alpha=0.015$  
(a) $t_{c1}=0.28$, $t_{c2}=0.08$, $t_{f1}=0.84$ and $t_{f2}=1.02$.
(b) $t_{c1}=0.08$, $t_{c2}=0.18$,$t_{c3}=0.31$,$t_{c4}=0.03$, 
$t_{f1}=0.72$,$t_{f2}=1.21$,$t_{f3}=2.00$ and $t_{f4}=0.15$.
}
\label{super}
\end{figure}

\end{document}